\documentclass[runningheads,a4paper]{llncs}

\usepackage{amssymb}
\usepackage{graphicx}
\usepackage{alltt}
\usepackage{url}
\urldef{\mailsa}\path|{michs,erwinl}@kth.se|
\newcommand{\keywords}[1]{\par\addvspace\baselineskip
\noindent\keywordname\enspace\ignorespaces#1}

\begin{document}

\mainmatter

\title{Performance Analysis of Irregular Collective Communication with the 
Crystal Router Algorithm}

\titlerunning{Performance Analysis of Irregular Collective Communication}

\author{Michael Schliephake \and Erwin Laure}
\authorrunning{Performance Analysis of Irregular Collective Communication}

\institute{KTH Royal Institute of
Technology\\CSC School of Computer Science and
Communication\\Department for High Performance Computing and
Visualization and\\ 
SeRC - Swedish e-Science Research Center,\\
SE-100 44 Stockholm, Sweden\\
\mailsa}

\toctitle{Lecture Notes in Computer Science}
\tocauthor{Authors' Instructions}
\maketitle

\begin{abstract} 
In order to achieve exascale performance it is important to detect
potential bottlenecks and identify strategies to overcome them. For this,
both applications and system software must be analysed and potentially
improved. The {\small EU FP7} project \emph{Collaborative Research into
Exascale Systemware, Tools \& Applications} {\small (CRESTA)} chose the
approach to co-design advanced simulation applications and system software
as well as development tools. In this paper, we present the results of a
co-design activity focused on the simulation code {\small NEK5000} that
aims at performance improvements of collective communication operations. We
have analysed the algorithms that form the core of {\small NEK5000'}s
communication module in order to assess its viability on recent computer
architectures before starting to improve its performance. Our results show
that the crystal router algorithm performs well in sparse, irregular collective
operations for medium and large processor number but improvements for even
larger system sizes of the future will be needed.
We sketch the needed improvements, which will make
the communication algorithms also beneficial for other applications that
need to implement latency-dominated communication schemes with short
messages. The latency-optimised communication operations will also become
used in a runtime-system providing dynamic load balancing, under
development within {\small CRESTA}.
\end{abstract}

\section{Introduction}

The development of applications showing exascale performance proves to be
very challenging. On one side, it comprises efforts to scale today's
numerical algorithms, system software, and development tools with proven
methods as well as the refactoring of non-optimal code pieces that would
become bottlenecks in runs at larger scale. On the other side, the
development of exascale applications includes the search for qualitatively
new approaches that reduce the computational complexity especially of
algorithms with non-linear scaling for increasing processor counts.
\emph{Collaborative Research into Exascale Systemware, Tools \&
Applications} {\small (CRESTA)} is an {\small EU FP7} project that
concentrates on the study and solution of issues that are connected with
the development towards exascale computing~\cite{CRESTA}. {\small CRESTA}
chose an approach based on the co-design of advanced simulation
applications and system software. The development of simulation codes has
been flanked with the further development of necessary developer tools like
parallel debuggers and performance analysis tools. {\small CRESTA'}s
co-design applications are running at the limits of available {\small HPC}
computer installations while researchers create an ever-increasing demand
for larger, respectively faster simulations and new application fields.
This tension provides requirements and challenges for system software and
tool developers. More demanding use cases can be used at the same time as
test cases of new developments and are checkpoints to assess improvements
though, for the time being still on current computers. Additionally, this
approach provides general lessons usable in more simulation applications
like those developed in the \emph{Swedish e-Science Research Centre}
{\small (SeRC)} as well as in the development of future software
development tools~\cite{SeRC}.

In this paper, we present the results of a co-design activity focused on
the simulation code {\small NEK5000} that aims at performance improvements
of collective communication operations. {\small NEK5000} can be used for
simulations of fluid flow, heat transfer and magnetohydrodynamics problems.
It is an open-source code mainly developed at the Mathematics and Computer
Science Division of the Argonne National Lab.

{\small NEK5000} is a mature solver for incompressible Navier-Stokes
equations. The numerical algorithm utilises high-order spatial
discretisation with spectral elements and high-order semi-implicit time
stepping for the calculations~\cite{Tufo1999}. An important property of the
algorithm is its fast convergence and the comparatively low complexity with
respect to the number of grid points \(n\). The complexity limits are for
the discretisation at \(O(n^6)\). The computational work and memory
accesses only require costs of \(O(n^4)\) and \(O(n^3)\)
respectively~\cite{Tufo2001}. The application has won the Gordon Bell Prize
in 1999 and many simulation projects on different {\small HPC} computer
installations show its scalability up to one million cores. Despite its
excellent scaling behaviour, the crystal router still exposes areas for
improvements. Our on-going co-design activity aims at implementations of
effective collective communication operations for large-scale runs as well
as the reduction of the communication volume using a hybrid parallelisation
scheme~\cite{Schliephake2012}.

In this paper, we present an analysis of the crystal router algorithm,
which is the base of {\small NEK5000'}s central communication module. It
will allow to use this solution as a base for the implementation of
alternative, improved collective communication operations. We identify
bottlenecks and sketch strategies to overcome these. These new collective
operations can be used also in other applications as well as a building
block in a runtime-system, which helps to dynamically improve load
balancing~\cite{Schliephake2011}. The remainder of this paper is organized
as follows: After a discussion of related work in Section 2 we describe the
functionality of the crystal router in Section 3. Benchmark results will be
presented and discussed in Section 4. Section 5 concludes the paper with an
outlook on future work.

\section{Related Work}

Sur et. al. developed efficient routines for personalized all-to-all
exchange on Infiniband clusters~\cite{Sur2004}. They use Infiniband RDMA
operations combined with hypercube algorithms and achieved speedup factors
of three for short messages of 32~B on 16 nodes.

Li et. al. use Infiniband's virtual lanes for the improvement of collective
MPI operations in multi-core clusters~\cite{Li2009}. These virtual lanes
are used for balancing multiple send requests active at the same time and
to increase the throughput for small messages. This implementation showed a
performance improvement of 10--20\thinspace\%.

Li et. al. analyse the influence of synchronisation messages on the
communication performance. Those messages are used in collective operations
to control of the exchange of large messages~\cite{Li2011}. They found that
contention of synchronisation messages accounts for a large portion of the
operation's overall latency. Their algorithm optimises the exchange and
achieved improvements of 25\% for messages between 32 and 64~kB length.

Tu et. al. propose a model of the memory-hierarchy in multi-core clusters
that uses horizontal and vertical levels~\cite{Tu2012}. Their experimental
results show that this model is capable to predict the communication costs
of collective operations more accurately than it ws possible before. They
developed a methodology to optimize collective operations and demonstrated
it with the implementation of a multi-core aware broadcast operation.

\section{Functionality of the Crystal Router}

The crystal router as developed by Fox et. al. \cite{Fox1988} is an
algorithm that allows sending messages of arbitrary length between
arbitrary nodes in a hypercube network. It is advantageous especially in
irregular applications where the exact nature of the communication is not
known before it occurs or where the message emergence changes dynamically.

Communication operations in hypercube networks are often implemented by
routing algorithms that iterate over the dimensions of the cube and execute
in each step one point-to-point communication operation with the partner
node at the other end of the respective edge. As explained for example in
\cite{Grama2003}, the result of the binary xor function with the processor
numbers of sender and receiver node as arguments provides a routing path
that can be used to transport the message. Therefore, messages can be
delivered in algorithms following this pattern from each node to each other
node in at most $d$ communication steps where $d$ is the dimensionality of
the hypercube network. In our implementation, we interpret MPI processes as
nodes of a hypercube network and use MPI ranks as processor numbers.

It has been proven that such a choice of paths provides load balancing in
the communication of several typical applications as well as it is optimal
if all processors are used in a load balanced way \cite{Grama2003}. The
crystal router has been developed to handle one typical situation of
processes in hypercube networks. In each process, there is a set of
messages, which must be sent to other processes. Destination processes
expect messages, but they know neither exactly how many messages will
arrive nor from which processes they will be sent. Nevertheless, the
communication happens for many algorithms typically in communication phases
between computations in a time-synchronised manner. One example is the
irregularity in the communication of molecular dynamics algorithms. The
real amount of data that has to be comunicated between neighbouring
subdomains is not known before the data exchange itself. Another example of
slightly irregular communication can be found in finite elelement
calculations where the meshes must be decomposed over several processors.
This decomposition will be perfect only to a certain degree. Therefore, the
communication between the nodes holding the different subdomains will show
some load-imbalance.

{\it Algorithm 1} explains how the transport of messages between arbitrary
processes works. First, all messages are stored in a buffer for outgoing
messages of the sender process ({\ttfamily msg\_out}). During the iteration
over the different channels (i.e. the bits of rank numbers), some messages
will be transmitted in each iteration step according to their routing path.
For that, those messages that must be transferred through a certain channel
will be copied from {\tt msg\_out} to a common transfer buffer ({\tt
msg\_next}). The buffer {\tt msg\_next} of each process will be exchanged
through the active channel of the current iteration step with the
respective buffer of a partner process. Thereafter, all messages that had
to be routed from this partner over this channel can be found in {\tt
msg\_next}. They will be inspected there. Messages that are addressed to
the receiving process will be copied into the buffer for incoming messages
({\tt msg\_in}) from where they can be accessed by the application code
later. Messages that have to be forwarded further in one of the following
iteration steps will be kept and put into {\tt msg\_out}.

\bigskip
\noindent
{\bf Algorithm 1} Pseudocode of the crystal router algorithm, adapted from
\cite{Fox1988}.
\begin{alltt}
{\bf begin} crystal_router
    {\bf declare buffer} msg_out;  /* buffer for messages to send    */
    {\bf declare buffer} msg_in;   /* buffer for received messages   */
    {\bf declare buffer} msg_next; /* buffer for messages to send    */
                             /* in the next communication step */
  
    {\bf for each} msg in msg_out {\bf do}
        {\bf if} dest_rank(msg) == myrank {\bf then}
            {\it copy msg into msg\_in;}
    {\bf end for}
    {\bf for each} dimension of the hypercube i = 0,...,d-1 {\bf do}
        {\bf for each} message msg in msg_out {\bf do}
            {\bf if} (dest_rank(msg)&myrank)\(^2\sp{i}\) {\bf then}
                {\it copy  msg into msg\_next;}
        {\bf end for}
        {\it exchange buffer msg\_next with process(rank == myrank\(^2\sp{i}\));}
        {\bf for each} message msg in msg_next {\bf do}
            {\bf if} dest_rank(msg) == myrank {\bf then}
                {\it copy msg into msg\_in;}
            {\bf if} {\it msg needs to be routed further} {\bf then}
                {\it copy msg into msg\_out;}
        {\bf end for}
    {\bf end for}
{\bf end} crystal_router
\end{alltt}
\noindent
\medskip

Summarizing, this algorithm guarantees message delivery between arbitrary
processes within $d$ steps where $d$ is the dimensionality of the hypercube
network. Furthermore, it maximises the message lengths for each
communication step by bundling messages that have a segment of their
routing paths in common, provided that the necessary buffers can be
allocated with a sufficient size.

\section{Performance Analysis of the Crystal Router}

%\paragraph{Communication benchmark.}

We developed a synthetic benchmark for the analysis of the original crystal
router algorithm. Its design has been based on the the communication
pattern in {\small NEK5000}. There, elements usually have 26 neighbour
elements. Each of them could be located in a different process, i.e.
processes have to exchange data with at least 26 neighbours due to spatial
domain decomposition. The element distribution logic tries to keep
neighbouring elements in processes on nodes near to each other, but, it is
also possible that some elements will be placed on distant nodes. Our
benchmark allows to define the number of communication partners of each
process as well as their distance in form of a stride that will be used to
select them. Selected nodes will exchange messages during the benchmark
run. The overall number of spectral elements per node, which corresponds to
a certain message length, could be adapted in order to test strong scaling.
In the strong scaling case, the volume-surface ratio of the elements located in one process causes a communication amount per node that is proportional to the number $p$ of processes with $O(p^{-2/3})$. The aggregated communication of the
job then follows the function $O(p^{1/3})$. The number of elements as well
as the amount of communication per process remains constant for weak
scaling. The aggregated communication of the parallel job will be
limited by $O(p)$ though.

%\paragraph{Measurements.}

The measurements have been done on KTH's system Lindgren. It is a Cray XE6
installation equipped with two AMD Opteron 6172 processors ("magny core")
and 32~GB RAM per compute node. It has a size of 1516 nodes, i.e. 36384
cores, and provides 305~TFLOPS peak performance. The system interconnect is
a Cray Gemini network with a 3D-torus topology~\cite{Alverson2010}.

\begin{figure}
\centering
\includegraphics[width=11cm]{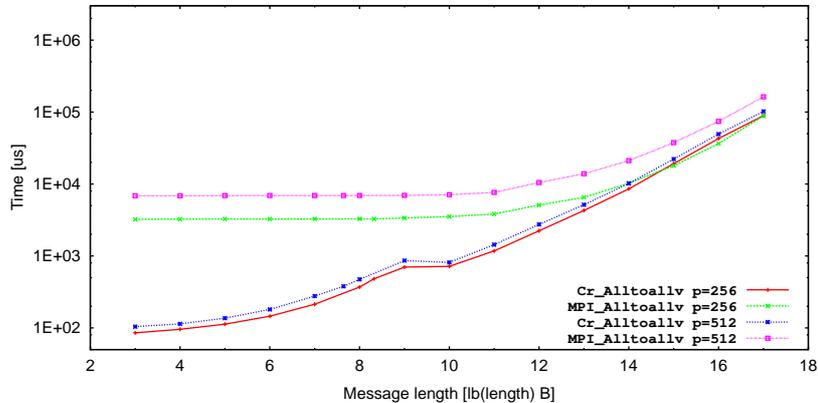}
\caption{Benchmark of personalized all-to-all communication implemented with
the crystal router based function {\tt Cr\_Alltoallv} and the MPI function
{\tt MPI\_Alltoallv}. Each process sends and receives data from 26 neighbouring
processes. The measurements have been executed with 256 respectively 512
processes.}
\label{fig:fig1a}
\end{figure}

\begin{figure}
\centering
\includegraphics[width=11cm]{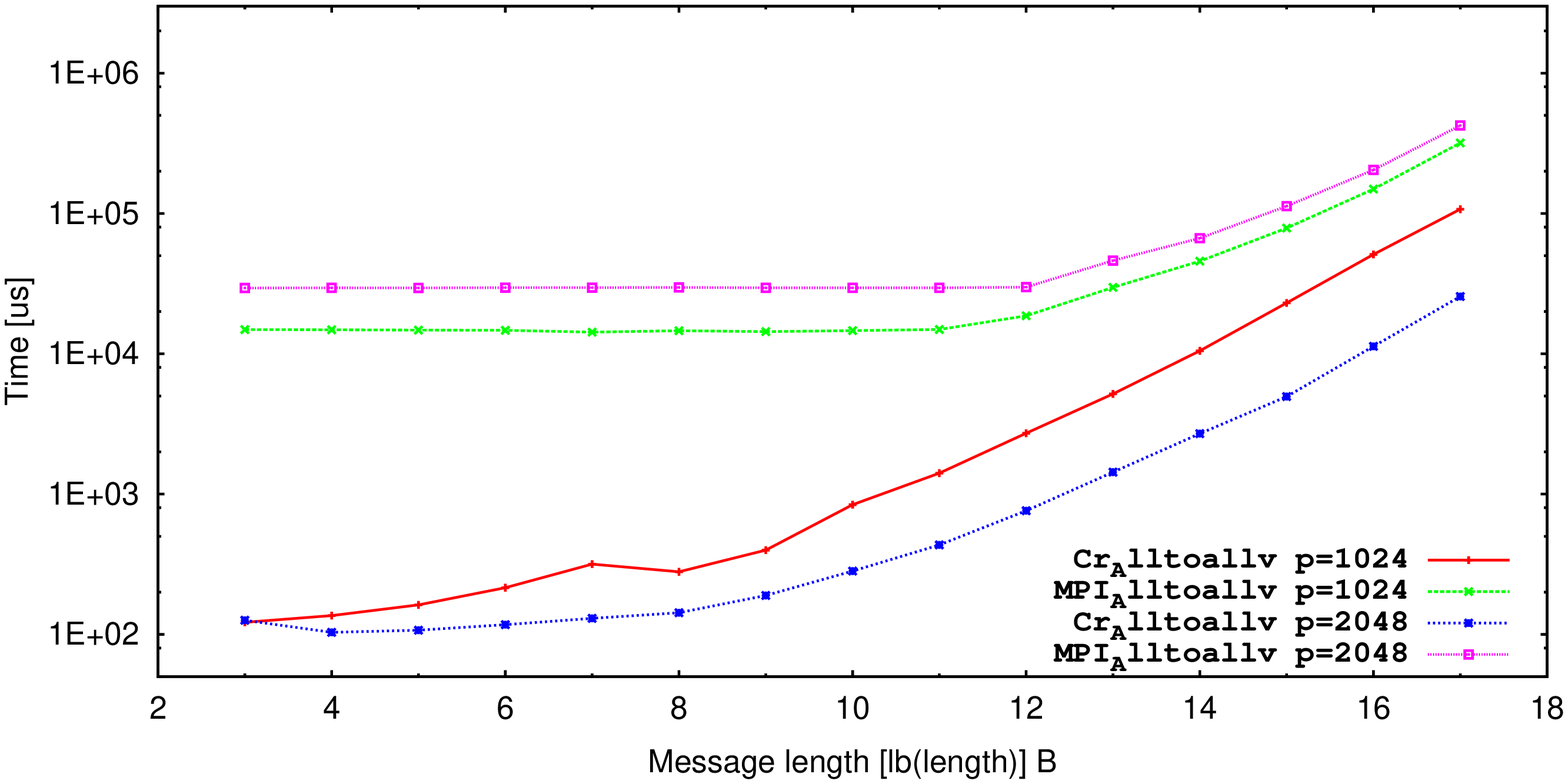}
\caption{Benchmark of personalized all-to-all communication implemented with the
crystal router based function {\tt Cr\_Alltoallv} and the MPI function
{\tt MPI\_Alltoallv}. Each process sends and receives data from 26
neighbouring processes. The measurements have been executed with 1024
respectively 2048 processes.}
\label{fig:fig1b}
\end{figure}

\begin{figure}
\centering
\includegraphics[width=11cm]{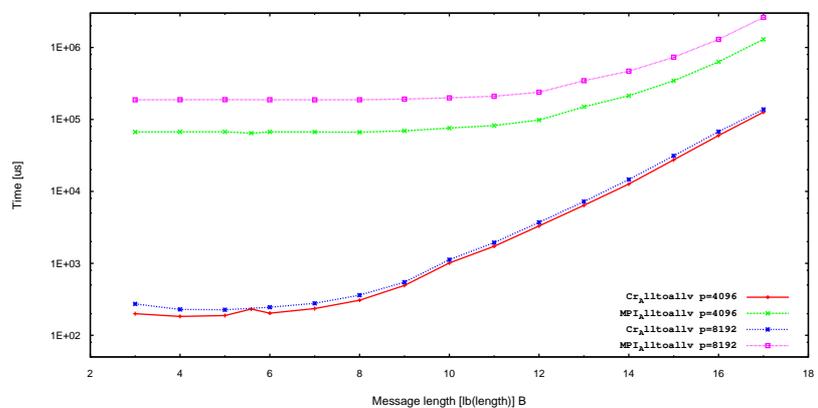}
\caption{Benchmark of personalized all-to-all communication implemented with the
crystal router based function {\tt Cr\_Alltoallv} and the MPI function
{\tt MPI\_Alltoallv}. Each process sends and receives data from 26
neighbouring processes. The measurements have been executed with 4096
respectively 8192 processes.}
\label{fig:fig1c}
\end{figure}

The first benchmark shows the performance of the crystal router for
different message lenghts and numbers of nodes in comparison to the
standard MPI library of Lindgren. The benchmarked operation is a
personalized all-to-all communication that is provided as {\tt
MPI\_Alltoallv}. The crystal router based implementation is called {\tt
Cr\_Alltoallv}. The benchmark has been setup in such a way, that each MPI
process communicates with its 26 nearest neighbours. The results for runs
with 256 and 512 processes are shown in Figure \ref{fig:fig1a}. The results
for 1024 and 2048 processes are shown in Figure \ref{fig:fig1b}. Finally, Figure \ref{fig:fig1c} provides results for 4096 and 8192 processes.

The crystal router based implementation {\tt Cr\_Alltoallv} is much faster than
{\tt MPI\_\allowbreak All\allowbreak to\allowbreak allv} in runs of all sizes especially for short, latency-bound messages. For example, $85\:\mu s$ are needed for a {\tt Cr\_Alltoallv} operation that lets each rank exchange 8 Bytes with its partner processes in a run with 256 processes. The operation takes $273\:\mu s$ for 8192 processes. The ratio of these times is $1 : 3.2$. The same operation needs $3\,227\:\mu s$ for 256 processes and $187\,000\:\mu s$ with 8192 processes with the function {\tt MPI\_Alltoallv}. The ratio of the times is $1 : 58$. This result demonstrates that sparse communication patterns involving all processes of a parallel program can be realised efficiently by the crysral router.

The speed advantage of the crystal router becomes smaller for longer messages. The speeds of the MPI system function and of the crystal router are almost equal for the longest messages of 128 kB in the smallest test of 256 processes. The speed difference increases for this message length with an increasing processor count and reaches a factor of 19 for the largest run utilising 8192 processes.

\begin{figure}
\centering
\includegraphics[width=11cm]{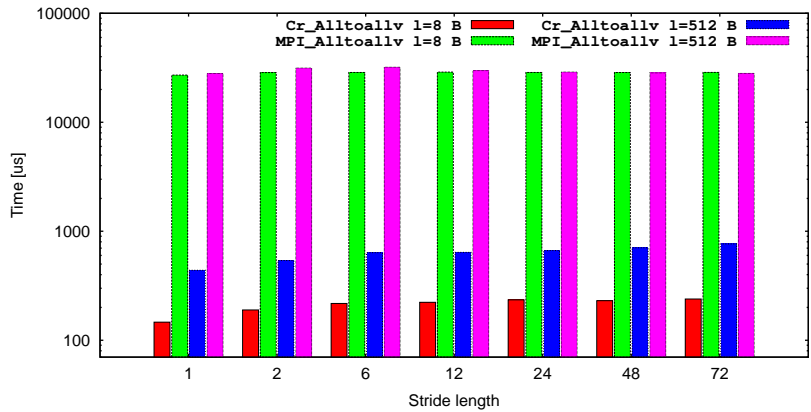}
\caption{Benchmark of all-to-all personalized communication as function of
the distance between communicating processes in the process list (stride).
The measurement has been executed with 256 processes.}
\label{fig:fig3}
\end{figure}

Furthermore, the benchmarks show that the number of communication partners
respectively the size of the stride do not noticeably influence the
duration of the operation for the MPI system function. The crystal router implementation contrastingly is more sensitive to these parameters.
Figure \ref{fig:fig3} shows measurements for a
varying stride length utilizing 2048 processes and transmitting messages of 8 resp. 512 byte length. The crystal router needs an increasing runtime for increasing strides. This reflects that the increasing stride length between the communications causes increasing data amounts that must be transfered the processes that are located on other numa nodes, on other sockets and on other nodes. For example, the time needed for the communication operation with a stride of 24 (i.e. each process communicates only with processes that reside on other nodes) is compared to a 1-stride 59\% longer for messages of 8 byte length, and it needs 51\% more time for messages of 512 byte length. Such a {\nobreak systematic} trend could not be observed with the MPI routine. Its variability is clearly smaller than 10\%.

\begin{figure}
\centering
\includegraphics[width=11cm]{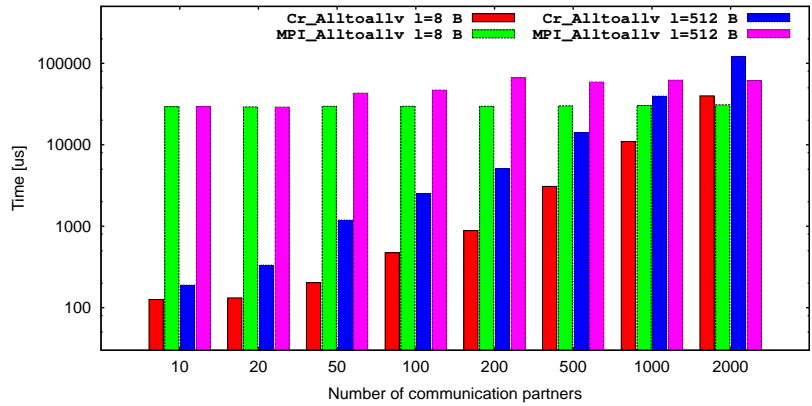}
\caption{Benchmark of all-to-all personalized communication as function of
the number of communication partners of each process. The measurement has
been executed with 256 processes.}
\label{fig:fig4}
\end{figure}

Figure \ref{fig:fig4} presents a benchmark that has been executed with
256 processes. Here, the number of communication partners of the processes
has been varied. The MPI system routine again does not show significant variations in their runtime. The crystal router implementation needs longer runtimes for an increasing number of communication partners per process. The result reflects the increasing communication volume that has to be processed by the constant number of processors.

\begin{figure}
\centering
\includegraphics[width=11cm]{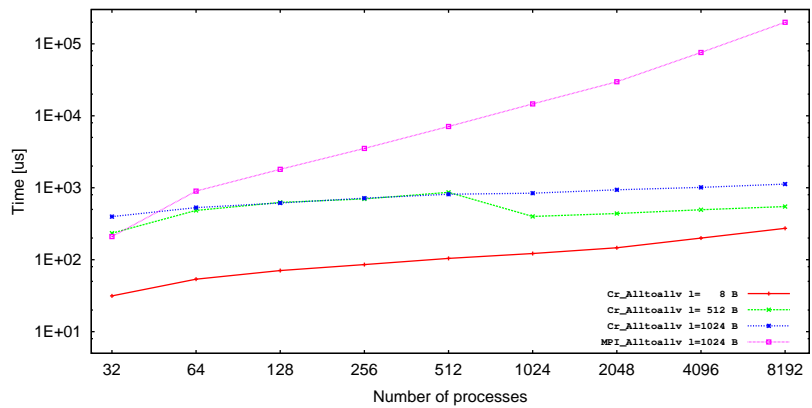}
\caption{Weak-scaling of all-to-all personalized communication with
{\tt Cr\_Alltoallv} for message lengths of 8, 512, and 1024 B. For
comparison the scaling of {\tt MPI\_Alltoallv} for a message length of
1024 B has been given.}
\label{fig:fig5}
\end{figure}

Finally, the evaluation with respect to weak scaling in Figure
\ref{fig:fig5} demonstrates that  {\tt Cr\_Alltoallv} scales very uniformly for short messages. Its scaling behaviour is noticeable better than that of {\tt MPI\_Alltoallv}.

Our analysis shows that the crystal router algorithm is beneficial for medium-sized and large parallel programs. It can unfold its capabilities compared to the  function {\tt MPI\_Alltoallv} especially in situations with sparse communication patterns and for large processor counts. Its uniform scaling into ranges of
large processor counts indicates that there are no effects of performance
degeneration in the algorithm itself and that it can be a viable choice for the implementation of collective communication operations. However, several improvements of the original algorithm are needed, particularly

\begin{itemize}
\item the reduction of data copies,
\item the exploitation of multiple communication paths, and
\item the overlapping of data transfers with the processing of the messages.
\end{itemize}

Specifically on Cray systems, the exploitation of multiple communication
paths and the overlapping of data transfers with the process-internal
message handling will provide significant performance improvements. The
3D-torus connects to each Gemini chip with several links. The Block
Transfer Engine {\small (BTE)} of the Gemini chip allows to offload the
transer of larger messages from the CPU. Therefore, a refatcoring of the
original algorithm using these hardware capabilities will extend the range
of its applicability.

\section{Conclusions and Future Work}

We evaluated the original crystal router algorithm in an implementation of
a personalized all-to-all communication on a recent computer architecture.
It shows a superior exchange performance especially for short messages up
to 4~kilobyte and
parallel runs of medium and large sizes. It showed furthermore a uniform
scaling over the whole range of job sizes. This is possible because it
bundles short messages into larger packages that will be transferred
at once. The influence of latency is reduced in that way, and MPI library
optimisations with respect to the bandwidth of larger message lengths
become useable for shorter messages too. The crystal router is sensitive slightly to the distance of the communicating processes and to a larger extend to the number of communication partners per process, i.e. the degree of sparsity. These comparatively small variations and the high overall efficiency that is achieved at the same time are an effect of the algorithm's properties. The
message bundling and the algorithm design guarantee the message
delivery within a fixed number of communication steps. Finally, the
hypercube algorithm involves all nodes equally into the transport of
messages during each communication step.

Our benchmarks confirm that the crystal router algorithm could be used
efficiently also on modern computer architectures, however, to make it
ready for exascale, the efficiency on higher processor counts needs to be
improved furthermore. We have sketched key aspects of these improvements, particularly
the reduction of data copying and the use of multiple network connections.
These improvements will make the cystal router based communication
substrate a viable choice for exascale applications.

\section*{Acknowledgements}

We would like to thank for the support of this work through the projects
\emph{Collaborative Research into Exascale Systemware, Tools \&
Applications} {\small (CRESTA)} and \emph{Swedish e-Science Research
Centre} {\small (SeRC)}.

\end{document}